%% file: main.tex
\documentclass[conference]{IEEEtran}
\IEEEoverridecommandlockouts
\usepackage{import}
\usepackage{cite}
\usepackage{amsmath,amssymb,amsfonts}
\usepackage[lined,boxed,vlined,ruled,linesnumbered]{algorithm2e}
\usepackage{algpseudocode}
\usepackage{graphicx}
\usepackage{textcomp}
\usepackage{xcolor}
\def\BibTeX{{\rm B\kern-.05em{\sc i\kern-.025em b}\kern-.08em
    T\kern-.1667em\lower.7ex\hbox{E}\kern-.125emX}}

\input{gmacros}

\begin{document}

\title{Distributed Learning for Proportional-Fair Resource Allocation in Coexisting WiFi Networks}

 \author{
    \IEEEauthorblockN{Piotr Gaw{\l}owicz\IEEEauthorrefmark{1}, Jean Walrand\IEEEauthorrefmark{2}, Adam Wolisz\IEEEauthorrefmark{1}}
    \IEEEauthorblockA{\IEEEauthorrefmark{1}Technische Universit\"at Berlin, Germany, \{gawlowicz, wolisz\}@tkn.tu-berlin.de}
    \IEEEauthorblockA{\IEEEauthorrefmark{2}University of California, Berkeley, USA, walrand@berkeley.edu}
}

\maketitle

\input{sections/abstract}


%
\input{sections/introduction}
%
\input{sections/relatedwork}
%
\input{sections/primer_wifi}

%
\input{sections/primer_sco}

%
\input{sections/design}
%
\input{sections/experiments_all}

%
\input{sections/conclusions}

\bibliographystyle{IEEEtran}
\bibliography{biblio,IEEEabrv}

\end{document}

%% file: gmacros.tex
\newcommand{\real}{\mathbb{R}}

\newcommand{\norm}[1]{\left\|#1\right\|}

\newcommand{\pa}[1]{\left(#1\right)}

\usepackage{todonotes}
\definecolor{PalePurp}{rgb}{0.66,0.57,0.66}



%% file: sections/abstract.tex

\begin{abstract}

In this paper, we revisit the widely known \textit{performance anomaly} that results in severe network utility degradation in WiFi networks when nodes use diverse modulation and coding schemes. 
The proportional-fair allocation was shown to mitigate this anomaly and provide a good throughput to the stations.
It can be achieved through the selection of contention window values based on the explicit solution of an optimization problem or, as proposed recently, by following a learning-based approach that uses a centralized gradient descent algorithm.
In this paper, we leverage our recent theoretical work on asynchronous distributed optimization and propose a simple algorithm that allows WiFi nodes to independently tune their contention window to achieve proportional fairness.
We compare the throughputs and air-time allocation that this algorithm achieves to those of the standard WiFi binary exponential back-off and show the improvements.

\end{abstract}

%% file: sections/introduction.tex

\section{Introduction}
The widely deployed WiFi technology uses Distributed Coordination Function (DCF),  a simple decentralized channel access mechanism based on the CSMA/CA protocol to share the limited wireless channel capacity. With DCF, the nodes adjust their channel access probability using a back-off mechanism. Specifically, they maintain a dynamically changing contention window (CW) and select the time lag between two consecutive transmission attempts as a random number within the scope of the CW.
By design, DCF ensures equal long term channel access probabilities to all involved stations, thus guaranteeing frame-level fairness.
However, the frame-fairness leads to the so-called \textit{performance anomaly} problem when stations use different transmission rates.  

Stations employ various modulation and coding schemes (MCSs) to preserve transmission robustness in response to the quality of the radio link they experience, which in turn depends on their communication distance, mobility, and other factors.
The lower MCSs offer increased robustness against errors at the cost of a drop in data rate.
Consequently, the WiFi stations that use lower data rates occupy the channel for a longer period of air-time that could be otherwise used more efficiently by the faster clients.
Precisely, under the frame-level fairness, a station with poor transmission conditions captures extensive air-time share and reduces the air time available to other stations, and hence the station \textit{slows down} all stations. 
This pathological behavior was first identified in 802.11b networks\cite{1208921}.
Then, Patras \textit{et al.}\cite{patras2016proportional} demonstrated that the effect is dramatically exacerbated in today's high throughput networks (i.e., 802.11n/ac), where the data rates among stations may vary by orders of magnitude, e.g., the throughput of a station using the data rate of 780\,Mbps become similar to that of a co-existing station with the data rate of 6\,Mbps.

The proportional-fair allocation, introduced by Kelly \mbox{\textit{et al.}}\cite{kelly1998proportional}, was shown to address the performance anomaly problem appropriately\cite{checco2011fairness}.
By definition, it maximizes the network utility (defined as the sum of the logarithms of individual throughputs) subject to the constraints that the communication conditions impose on the individual stations. 
In~\cite{5370273}, the proportional fairness allocation in WiFi networks was formulated as a convex optimization problem that is solved by selecting optimal contention window values for the stations.

The existing approaches to solve this problem are deployed in a central node (e.g., AP or a controller node) and require knowledge of the average frame duration or throughput of each transmitting station.
However, such centralized operation cannot be assumed in the case of overlapping but separately managed WiFi networks. Furthermore, the standard does not envision the possibility of communication between nodes belonging to different networks.

Consequently, in this paper, we propose a distributed learning-based approach where WiFi nodes independently tune their own contention window to achieve proportional fairness.
Our algorithm is based on a stochastic convex optimization framework. 
Specifically, it is based on our previous work~\cite{walrand2020distributed} that proves the convergence of the Kiefer-Wolfowitz algorithm~\cite{kiefer1952stochastic} in a distributed and asynchronous setting. 
Those properties are highly beneficial for coexisting WiFi nodes as they allow learning channel access parameters without any coordination nor explicit information exchange. 
Instead, the nodes use overheard frames to compute the network utility and independently follow a gradient descent method to optimize overall network performance.
In general, we explore the usability of the distributed and asynchronous Kiefer-Wolfowitz (DA-KW) algorithm in a practical use-case of wireless optimization. 
Specifically, the contributions of this work are as follow:
\begin{itemize}
    \item We propose a simple approach for distributed contention window tuning that achieves proportional fairness in coexisting WiFi scenarios. To this end, we apply the DA-KW algorithm to the WiFi domain while introducing slight modifications to address the practical issues.
    \item Using simulations, we evaluate the proposed approach in terms of convergence speed and achieved performance in multiple scenarios. Moreover, we compare its performance with the standard WiFi DCF. 
    \item We investigate the impact of the level of coordination (i.e., synchronized execution). It appears that there is no significant gain of the coordination in the case of a single collision domain. However, coordination and information exchange allow achieving optimal channel allocation in the case of overlapping collision domains.
\end{itemize}

%% file: sections/relatedwork.tex

\section{Related Work} \label{chapter:related_work}

Proportional fairness in WiFi networks has been extensively studied, e.g.,~\cite{6195728, 6598666, 5463215}.
Checco \textit{et al.}~\cite{checco2011fairness} provided an analysis of proportional fairness in WiFi networks. 
The authors proved that a unique proportional fair rate allocation exists and assigns equal total (i.e., spent on both colliding and successful transmission) air-time to nodes.
Patras \textit{et al.}~\cite{patras2016proportional} extended this analysis to multi-rate networks, and confirmed that under proportional fair solution 
all the stations get an equal share of the air-time, which is inversely proportional to the number of active stations.
The authors formulated network utility maximization as a convex optimization problem and provided a closed-form solution that can be solved explicitly.
To this end, a WiFi access point (AP) estimates the average frame transmission duration for each station and pass it as an input to an optimization tool. The computed CW values are distributed to nodes in a beacon frame. The optimization is executed periodically (i.e., every beacon interval) to react to changes in the network (e.g., traffic or wireless conditions).

An alternative approach was proposed by Famitafreshi \mbox{\textit{et at.}}~\cite{Golshan2020bco}. The authors use a stochastic gradient descent (SGD) algorithm, which can iteratively learn the optimal contention window only by monitoring the network's throughput.
Specifically, the learning agent resides in a WiFi AP, where it measures the uplink throughput of each connected station and send the $CW$ value updates to all the stations in a beacon frame. The algorithm combines two utility values measured under different $CW$ values to compute the gradient and update the $CW$ following the SGD algorithm.

Both proposed algorithms use global knowledge and centralized operation (i.e., deployed in AP). However, in typical scenarios, multiple networks under separate management domains are co-located and have to coexist, and there is no central entity with the full knowledge to perform the optimization or learning. 

%% file: sections/primer_wifi.tex


\section{Relevant WiFi Background}

In this section, we briefly describe the CSMA operation, present its analytical models, and summarize the throughput optimization in WiFi networks.

\subsection{WiFi Random Back-off Operation}

WiFi nodes use the DCF mechanism to access the channel. The DCF is based on the Carrier Sense Multiple Access/Collision Avoidance (CSMA/CA) method and employs binary exponential back-off (BEB) to control the contention window. 
Specifically, in DCF, before a frame transmission, a WiFi station has to perform a random back-off procedure. 
To this end, it initializes its back-off counter with a random number drawn uniformly from $\{0, \ldots, CW-1\}$, where $CW$ is the contention window.
Then, the station observes the wireless channel and decrements the counter whenever it is sensed idle during a DCF inter-frame space (DIFS) and freezes the back-off counter otherwise. 
Finally, when the back-off counter reaches zero, the station transmits a frame. If the transmission is successful (as indicated by the reception of an acknowledgment), the station sets $CW$ to the minimal value, i.e., $CW_{min}$, for the next transmission. Otherwise, it doubles the previous contention window and performs the frame retransmission. The $CW$ is increased until it reaches the maximal value defined by $CW_{max}$.

\subsection{Analytical Model of Contention-based Medium Access}

Here, we briefly describe the analytical model derived in~\cite{patras2016proportional} which allows computing the total throughput achieved by WiFi nodes.
We assume that all nodes are in a single collision domain (i.e., each node overhears transmissions of all other nodes). 
For the sake of clarity of presentation, in this section, we consider the case where all stations are saturated (i.e., they always have packets to transmit), but in Section VI, we also evaluate our algorithm in scenarios with non-saturated traffic.
Note that the model allows for arbitrary packet sizes and selection of MCS.

Let us consider a set of $N$ wireless stations, where each active station $i$ accesses the channel with slot transmission probability $\lambda_i$. 
The relation between channel access probability to a constant contention window is $CW_i = \frac{2-\lambda _i}{\lambda_i}$~\cite{bianchi2000performance}.
The transmission failure probability experienced by a station $i$ equals $p_{f,i} = 1 - (1 - p_{n,i})(1 - p_i)$, where $p_{n,i}$ is the probability that the transmission fails due to channel errors (e.g., noise or interference), while $p_{i} = 1 -  \prod_{j=1,j \neq i}^{N}(1-\lambda_j)$ denotes the collision probability experienced by a packet transmitted by this station.
Then, the throughput of station $i$ equals $S_i$:

\begin{equation}
\label{eg:si}
    S_i=\frac{p_{s,i}D_i}{P_e T_e  + P_s T_s + P_u T_u}
\end{equation}

\noindent where, $p_{s,i} = \lambda_i (1-p_{f,i})$ is the probability of a successful transmission performed by station $i$, while $D_i$ denotes its frame payload size in bits. 
$P_e = \prod_{i=1}^{N}(1-\lambda_i)$ is the probability that the channel is idle during a slot of duration $T_e$ (e.g., $9\mu s$ in 802.11n). 
$P_s = \sum_{i=1}^{N}p_{s,i}$ and $P_u = 1 - P_e - P_s$ are the expected probabilities of successful and unsuccessful transmission with the expected durations $T_s = \prod_{i=1}^{N} \frac{p_{s,i}}{P_s} T_{s,i}$ and $T_u = \prod_{i=1}^{N} \frac{p_{u,i}}{P_u} T_{u,i}$, respectively. 
Here, $T_{s,i}$ and $T_{u,i}$ are the durations of a successful and a failed transmissions of each station that depend on the fixed preamble duration, the variable duration of a header, the size of a payload transmitted with the PHY rate $C_{i}$, and whether an acknowledgment is sent (success) or not (failure).
Finally, $p_{u,i} = \tau_i p_{n,i} \prod_{j=1,j \neq i}^{N} \left ( 1 - \tau_j \right ) + \tau_i \left ( 1 -   \prod_{j=1}^{i-1} ( 1 - \tau_j ) \right ) \prod_{j=i+1}^{N} ( 1 - \tau_j )$ is the probability of an unsuccessful transmission (either due to collision or channel errors) of stations of highest index $i$ (when labeling stations according to their transmission durations). 
Note that the proper labeling is needed as the duration of a collision is dominated by the frame with the longest duration involved in that collision, and collisions should only be counted once.

Using the transformed variable $y_i=\frac{\lambda_i}{1-\lambda_i}$, the expression of a station’s throughput (\ref{eg:si}) can be rewritten as:
\begin{equation}
\label{eg:si2}
    S_i=(1-p_{n,i})\frac{y_i}{Y} D_i
\end{equation}
\noindent where $Y = T_e + \sum_{i=1}^{N}\left ( y_i T_{s,i} \prod_{k=1}^{i-1} \left ( 1 + y_k \right )\right )$. 

We refer to~\cite{patras2016proportional} for the details of the model and the transition from identity (\ref{eg:si}) to (\ref{eg:si2}).

\subsection{Proportional-fair Allocation}

Following \cite{checco2011fairness}, we formulate the proportional-fair allocation problem as a convex optimization problem. 
The global utility function is defined as the sum of the logarithms of individual throughputs, i.e., $U = \sum_{i=1}^{N} \tilde{S_i}$, where $\tilde{S_i} = log(S_i)$. 
The utility maximization problem is as follows:

\begin{align*}
\begin{split}
    & \textup{maximize} \: \: \sum_{i=1}^{N} \tilde{S_i} \\
    & \textup{s.t.} \: \: \tilde{S_i} \leq log \left ( z_i \frac{y_i}{Y} D_i \right ), \: \: i = 1,2,...,N \\
    & \textup{and }  0 \leq y_i, \: \: i = 1,2,...,N \: \: (0 \leq \lambda_i \leq 1)
 \end{split} 
\end{align*}

\noindent where $z_i = 1-p_{n,i}$. The constraints ensure that the optimal solution is feasible, i.e., it is within the log-transformed rate region $\tilde{R}$. 
The rate region $R$ is a set of achievable throughput vectors $S(\mathbf{\lambda}) = \left [ S_1, S_2,...,S_N \right ]$ as the vector $\lambda$ of attempt probabilities ranges over domain $[0, 1]^N$.
The set $R$ is known to be non-convex in 802.11 networks, but the log-transformed rate region $\tilde{R}$ is strictly convex~\cite{5370273}. Moreover, as the strong duality and the KKT (Karush-Kuhn-Tucker) conditions are satisfied, a global and unique solution exists.

%% file: sections/primer_sco.tex

\section{Distributed Stochastic Optimization Primer}

Here, we briefly introduce the stochastic convex optimization techniques in centralized (i.e., single agent) and distributed (i.e., multiple agents) settings.

\subsection{Stochastic Convex Optimization}\label{sco}

A Stochastic Convex Optimization deals with minimizing of the expected value of a function $F(\mathbf{x}, \xi)$ that is convex in $\mathbf{x} \in \real^d$ where $\xi$ is random vector:
\begin{equation}
    \mbox{Find} \:  \mathbf{x}^* = \underset{\mathbf{x} \in \mathcal{K}}{\mbox{argmin}} \: f(\mathbf{x}) := E(F(\mathbf{x},\xi)).
\end{equation}
The setup is that one has access to sample values of $F(\mathbf{x}, \xi)$.  

If one could measure the gradient $\nabla f(\mathbf{x})$ of the function, one could use a gradient descent algorithm where the parameters at the $t$-th iteration, $\mathbf{x}_t$, are updated according to ${\mathbf{x}}_{t+1} = \mathbf{x}_{t} - \eta_t \nabla f(\mathbf{x})$, where $\eta_t$ is the step size at iteration $t$. 
However, in practical scenarios there is no access to the actual gradient $ \nabla f( \mathbf{x} )$. 
Accordingly, the stochastic gradient descent algorithms rely on constructing noisy gradient estimates $\tilde{g_t}$, which are then used to adjust the parameters according to $\mathbf{x}_{t+1} = \mathbf{x}_{t} - \eta_t \tilde{g_t}$.

The Kiefer-Wolfowitz (KW) algorithm~\cite{kiefer1952stochastic}  is a gradient estimation method that combines two function evaluations with perturbed values of its variable to compute the estimate. 
The simultaneous perturbation stochastic approximation (SPSA) algorithm~\cite{spall1992spsa} is an extension of the KW algorithm towards multivariate problems. In SPSA, the partial derivatives with respect to the different variables are estimated by simultaneously perturbing each variable by an independent and zero-mean amount, instead of perturbing the variables one at a time.

The optimization procedure can be performed in a centralized or distributed setting. In the former case, a single agent knows and controls all the variables in vector $\mathbf{x}$ and has the exclusive right to query the function (we refer to it as \textit{environment}), while in the latter case, those assumptions do not hold.

\subsection{Distributed Convex Optimization}\label{dco}

\begin{figure}[ht!]
\centering
\vspace{-5pt}
  \includegraphics[width=0.75\linewidth]{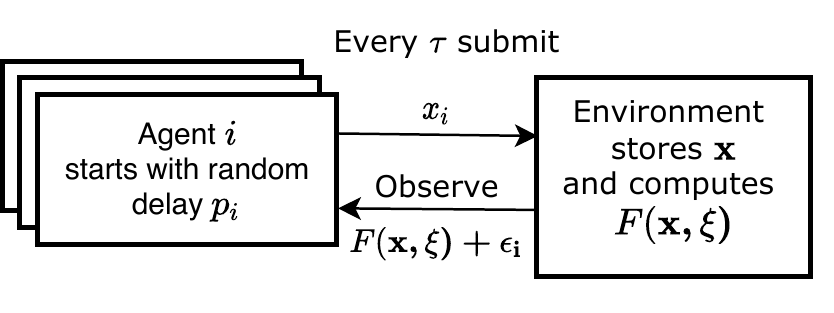}
  \vspace{-12pt}
  \caption{The interaction between distributed agents and the environment -- agents asynchronously submit their individual variables to the environment and get noisy observations of its value.} 
  \label{da_spsa}
  \vspace{-5pt}
\end{figure}

In the distributed settings, a set of $N$ distributed agents try to optimize a global objective function. The critical challenge is that agents make individual decisions simultaneously, and the value of the function depends on all agents' actions.
Moreover, we assume that the agents are not synchronized (i.e., they query the function and update their variable asynchronously at a random point in time) and cannot communicate.
Specifically, each agent adjusts his own variable $x_i$ without knowing the values of the other variables, i.e., $x_{-i}$. Fig.~\ref{da_spsa} shows the interaction between agents and the environment.

\begin{algorithm}[b]
\SetAlgoLined
{\bf Input:} Non-increasing sequences $\pa{\delta_k}, \pa{\eta_k}$ \\
{\bf Input:} Function sample interval $\tau \geq 1$ \\
{\bf Input:} Phase offset $p_i \in \{1, \ldots, \tau - 1 \}$ \\
{\bf Initialization:} Choose arbitrary $y_0 \in \mathbb{R}$ \\
  \For{$k=0,1,\dots,$}{
  Draw $\varepsilon_k$ uniformly from $\left\{-1,1 \right \}$ \\ 
  Let $t= k\times 2\tau + p_i$ \\
  At $t$ set $x_t = y_k + \varepsilon_k \delta_k$ \\ 
  Observe $g^+_k = f(\bf{x}_t)$. \\
  At $t + \tau$ set $x_{t+\tau} = y_k - \varepsilon_k \delta_k$ \\
  Observe $g^-_k = f(\bf{x}_{t+\tau})$ \\
  Compute gradient estimate $\tilde{g_k} = \frac{g^+_k - g^-_k}{2\varepsilon_k \delta_k}$ \\
  Update $y_{k+1} = y_k - \eta_k \tilde{g_k}$ \\
  \textbf{if} $\norm{x_t - x^{*}} < \varepsilon^{*}$ \textbf{break} 
  }
 \caption{DA-KW (executed by each agent $i$)}
 \label{da-dkw}
\end{algorithm}

Following a naive application of the KW algorithm, each agent estimates the gradient by perturbing its own variable by a zero-mean change and querying the global function every time interval $\tau$. Then, it uses those measurements to determine the gradient and updates its variable in proportion to the computed estimate.
Note that agents perform such experiments without any coordination. Thus, when an agent attempts to get a second evaluation of the function, the function may have already changed due to another agent's query. 
Intuitively, each agent gets gradient estimates corrupted due to the actions of other agents.
However, we have recently proved that the KW algorithm can converge to the optimal solution in a distributed and asynchronous (DA) setting if the size of perturbation is bounded~\cite{walrand2020distributed}. 
The algorithm executed by each agent is presented as Algorithm~\ref{da-dkw}.

%% file: sections/design.tex

\section{Distributed Contention Window Learning}

Based on a distributed and asynchronous Kiefer-Wolfowitz algorithm, we propose a simple collaborative learning scheme for WiFi nodes, which allows them to learn the optimal contention window values without coordination and information exchange.

\subsection{Practical Issues}

There are several practical issues that have to be considered.
First, the utility function evaluation is not immediate, as an agent cannot measure the instantaneous throughput. Instead, it has to count an amount of successfully transmitted data by overhearing frames and compute the mean throughput of neighboring nodes over the measurement time slot $\tau$.
As agents follow this procedure simultaneously with random phase offsets, the utility that each agent observes is an average of multiple function evaluations during $\tau$ that correspond to changing $CW$ of agents. 
Second, in contrast to the formal proof where the function is globally defined, the congestion window takes values in a discrete set between $CW_{min}$ and $CW_{max}$.
Finally, to make an algorithm responsive to changes (e.g., used data rates), we use constant exploration and learning parameters and remove the termination condition. Therefore, the algorithm cannot converge to the optimum value but only to its neighborhood.

\subsection{Proposed Approach}

We apply a modified version of the DA-KW algorithm to learn the IEEE 802.11's $CW$ cooperatively. More specifically, as illustrated in Algorithm~\ref{da-spsa-c}, we introduce modifications to match the discussed practical issues.

From the point of view of a single agent, our proposed technique works as follows. 
During initialization, a WiFi node selects a contention window value within the range of $\left [ CW_{min}, CW_{max} \right ]$.
The integer $CW$ value is converted to log-transformed variable $y$ using function $L$. Specifically, $L$ computes channel access probability as $\lambda = \frac{2}{CW + 1}$, and then transformed variable as $y=log (\frac{\lambda}{1-\lambda})$.
By $L^{-1}$, we designate the operation inverse to $L$. 

At a random time point $t$, node $i$ perturbs its log-transformed variable by a fixed exploration parameter $\delta_k$, i.e., it replaces $y_k$ by $y_k + \varepsilon_k \delta_k$ and converts that value back to the discrete $CW_t$ value. Next, for the duration of a single measurement slot $\tau$ (e.g., 100\,ms), it transmits all its frames applying the $CW_t$ value to the back-off procedure. 
Simultaneously, the node observes the environment formed by all WiFi nodes. That is, by overhearing frames, it counts the amount of data transmitted by each neighbor. At the end of the measurement slot, it computes the value of the network utility function, i.e., the sum of the logarithms of the observed throughputs of the different nodes and the known own throughput.
Then at $t+\tau$, the node again perturbs its log-transformed variable, i.e., it replaces $y_k$ by $y_k - \varepsilon_k \delta_k$, and repeats the measurement procedure. 
Finally, at $t+2\tau$, it combines both measurements to compute the gradient estimate and updates its log-transformed variable $y_{k+1}$ accordingly.
The value is projected to the decision set defined as $\mathcal{K}_{\alpha} = [A+\alpha,B-\alpha]$, where $\alpha \leq \frac{B-A}{2}$.
The projection of $x$ to the nonempty interval $[a, b]$ is defined as $\Pi_{[a,b]} = \mbox{max}\{\mbox{min}\{b, x\} , a\}$. 
Note that the gradient descent is performed in a continuous domain using log-transformed variable $y$, which is then converted and discretized into an integer $CW$ value. 

\begin{algorithm}[t!]
\SetAlgoLined
{\bf Input:} $L$ converts $CW$ to log-transformed value $y$\\
{\bf Input:} Constant parameters $\delta_k = \delta, \eta_k = \eta$ \\
{\bf Input:} Function sample interval $\tau$ \\
{\bf Input:} Random phase offset $p_i \in \left [ 0, \tau \right ]$ \\
{\bf Initialization:} Choose $y_0 \in \left [L(CW_{min}), L(CW_{max}) \right ]$ \\
  \For{$k=0,1,\dots,$}{
  Draw $\varepsilon_k$ uniformly from $\left\{-1,1 \right \}$ \\ 
  Let $t= k\times 2\tau + p_i$ \\
  At $t$ set $CW_t = \left \lceil L^{-1}(y_k + \varepsilon_k \delta_k) \right \rceil$ \\ 
  Observe transmissions of neighbors for $\tau$ and compute $g^+_k = \sum_{i=1}^{N} \tilde{S_i}$. \\
  At $t + \tau$ set $CW_{t+\tau} = \left \lceil L^{-1}(y_k - \varepsilon_k \delta_k) \right \rceil$ \\
  Observe transmissions of neighbors for $\tau$ and compute $g^-_k = \sum_{i=1}^{N} \tilde{S_i}$. \\
  Compute gradient estimate $\tilde{g_k} = \frac{g^+_k - g^-_k}{2\varepsilon_k \delta_k}$ \\
  Update $y_{k+1} = \Pi_{k} \left ( y_k - \eta_k \tilde{g_k} \right )$,   where $\Pi_k$ is the projection operator onto 
  $\mathcal{K}_{\delta_k}$}
 \caption{Proposed Algorithm (executed by each agent $i$)}
 \label{da-spsa-c}
\end{algorithm}

\subsection{Impact of Diverse Coordination Levels}

To evaluate the impact of action synchronization among the distributed nodes, we target the distributed scenario with a multi-step approach that gradually removes the coordination between nodes:

\noindent \textbf{Coordinated Learning: } 
In the case of full coordination, the measurement slots of agents are synchronized, i.e., $p_i = 0$ for $i \in \{1,..,N\}$. Specifically, agents perform the gradient estimation procedure and update the $CW$ at the same time.
Therefore, the utility function is evaluated with constant variables in each time slot.

\smallskip
\noindent \textbf{Slotted Learning:} In the second step, we remove the stage coordination between agents, i.e., the time is still slotted, and agents enter the gradient estimation procedure at the slot boundaries. However, they might be at a different stage of the procedure as $p_i \in \{0, \tau \}$ for $i \in \{1,..,N\}$. As a result, they perform a $CW$ update at two different time points.
The environment state does not change during a single measurement period. However, it changes in each slot. 
Hence, agents perform the first measurement (i.e., $g^-_k $) under a different environment state than the second one (i.e., $g^+_k $).

\smallskip
\noindent \textbf{Uncoordinated Learning:} 
In the general case, the actions of distributed agents are not synchronized, and at any point in time $t$ each agent is at a different stage of the algorithm. 
Note that from each agent's perspective, the environment state could change $N-1$ times within a single measurement period.

%% file: sections/experiments_all.tex

\section{Performance Evaluation}\label{sec:eval}

We evaluate the distributed contention window tuning algorithm by means of simulations using the ns-3 network simulator~\cite{ns3} in conjunction with ns3-gym~\cite{ns3gym}. 
Specifically, we use the model for IEEE 802.11n in infrastructure-based mode and create multiple overlapping WiFi networks. Note that nodes belonging to separate networks cannot communicate.
If not stated otherwise, we create a fully-connected topology (i.e., single collision domain), where nodes are uniformly spread in the area of 10-by-10\,m; hence, every transmitter can sense ongoing transmissions.
In order to change network contention conditions, we vary the number of transmitting nodes. Moreover, we change the data rates and frame sizes to influence the solution of the proportional-fairness problem.
To turn off the BEB procedure and enable simple uniform back-off, we assign the same value of $CW$ to $CW_{min}$ and $CW_{max}$. We bound the $CW$ value to the range used in WiFi, i.e., $CW \in \{15,1023\}$, hence the log-transformed $CW$ operates in $y \in \{-6.23, -1.94\}$.
We show the convergence of the distributed algorithm using the evolution of the contention window, air-time share, and throughput. We compare the performance of our technique against the original IEEE 802.11 BEB technique.

\subsection{Selection of Measurement Slot Duration}\label{sec:tau}

First, we evaluate the impact of the measurement slot duration on the quality of the network utility estimates and the algorithm's convergence. 

\begin{figure}[hb!]
  \begin{minipage}[b]{1.0\linewidth}
    \includegraphics[width=\linewidth]{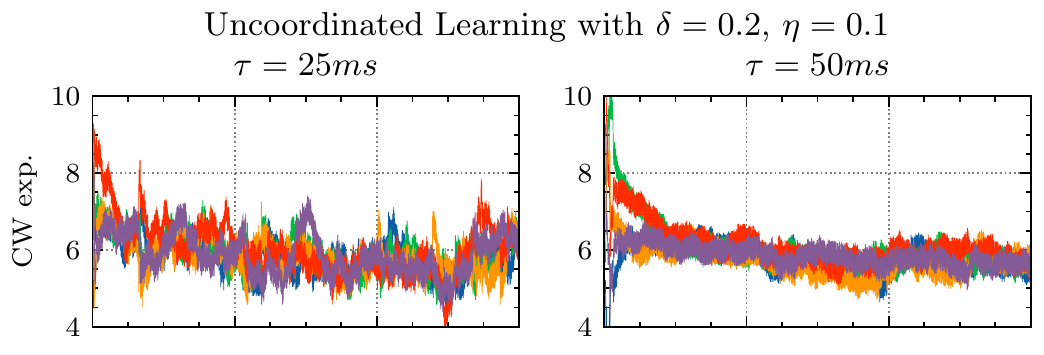}
  \end{minipage}\hfill
  \begin{minipage}[b]{1.0\linewidth}
    \includegraphics[width=\linewidth]{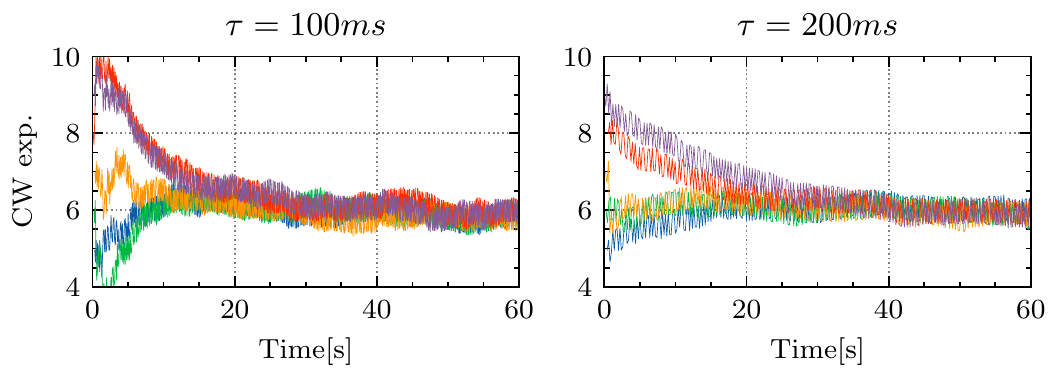}
  \end{minipage}\hfill
  \vspace{-5pt}
  \caption{Convergence of individual $CW$ under various measurement slot duration $\tau \in \{25, 50, 100, 200\}$\,ms.} 
  \label{fig:tau}
\end{figure}

To this end, we consider a scenario with five transmitting stations with homogeneous traffic, i.e., each station is backlogged with 1000\,B UDP packets and transmits to its own AP with a data rate equal to 26\,Mbps (MCS 3). 

Fig.~\ref{fig:tau} shows the evolution of the contention window value (we show $c = \log_2(CW)$) for each of transmitting nodes with four slot durations, $\tau \in \{25, 50, 100, 200\}$\,ms.
We observe higher variability for smaller values of $\tau$, which is expected due to the random nature of the frame transmissions. Specifically, with smaller values of $\tau$, the nodes cannot collect enough statistics to accurately estimate the network utility. 
These effects are alleviated by increasing the duration of $\tau$ so that the throughput estimates become more accurate, but at the cost of longer convergence time because updates are less frequent.
For further evaluation, we select $\tau = 200$\,ms, and leave its optimization as future work.

\begin{figure*}[ht!]
\centering
  \includegraphics[width=1.0\linewidth]{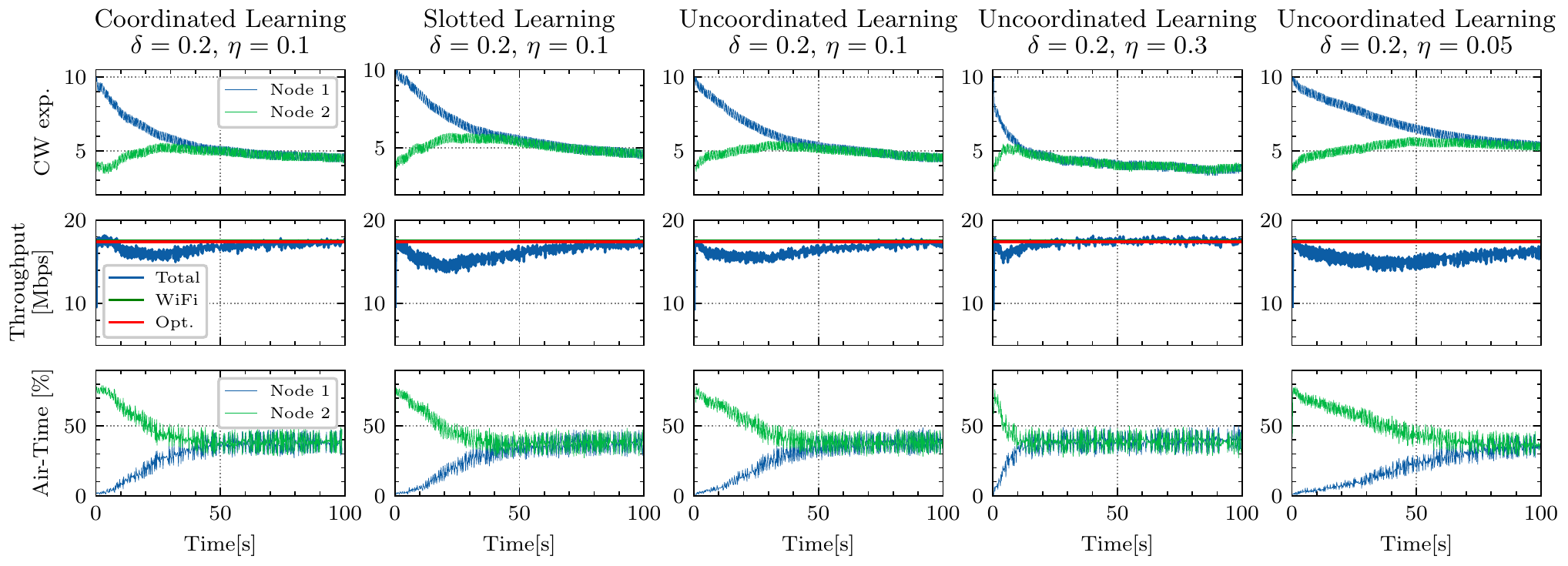}
  \vspace{-20pt}
  \caption{Convergence of the proposed algorithm with two transmitting nodes.}
  \label{fig:fc_2_comparison}
\end{figure*}

\begin{figure*}[ht!]
\centering
  \includegraphics[width=1.0\linewidth]{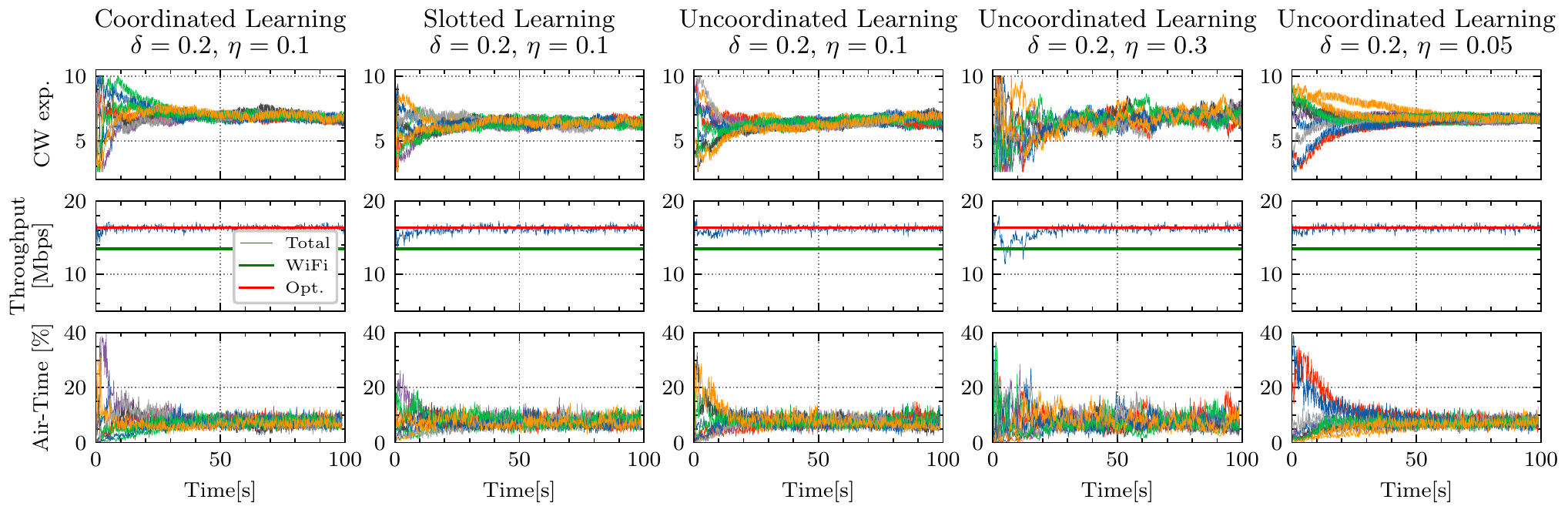}
  \vspace{-20pt}
  \caption{Convergence of the proposed algorithm with ten transmitting nodes.}
  \label{fig:fc_10_comparison}
  \vspace{-12pt}
\end{figure*}

\subsection{Homogeneous Traffic}
Here, we examine the sensitivity of the proposed distributed algorithm to the type of coordination and learning parameters. Moreover, we vary the number of active nodes with $N \in \{2,10\}$. 
We use the same homogeneous traffic parameters as in section~\ref{sec:tau}. 
Note that the exploration parameter $\delta$ and the learning rate $\eta$ were adequately selected to provide good performance.

Fig.~\ref{fig:fc_2_comparison} and Fig.~\ref{fig:fc_10_comparison} show a representative evolution of individual contention windows, total network throughput, as well as allocation of air-time in the scenario with two and ten nodes, respectively. 
In each setting, the nodes start with the same $CW$ values. In the case of two nodes, one starts with $CW = 1023$ and the second one with $CW = 15$. While, in the case of ten nodes, each starts with $CW$ value equal to $2^c$, where $c \in \left [4, 10\right]$.
First, we observe that the distributed nodes converge to similar $CW$ values, as expected because of the homogeneous traffic, and the algorithm converges to equal air-time allocation and optimal total network throughput. 
Note that in the case of ten active nodes, the total throughput is around 20\% higher than that achieved by WiFi.
The algorithm converges faster in the case of ten nodes, as the utility function becomes steeper with an increased number of nodes~\cite{Golshan2020bco}. The variability of $CW$ is higher for ten nodes, as the node estimates become noisier. Nevertheless, the network operates with approximately optimal utility.
Moreover, in the case of two nodes, we observe an interesting cooperative behavior where the initially more aggressive node slows down to free more air-time for its peer, then both nodes increase their aggressiveness to maximize the network utility.

Our results also show that the distributed algorithm behaves similarly with and without learning coordination, i.e., there is no significant advantage to coordination.

Finally, the three right-most columns in Fig.~\ref{fig:fc_2_comparison} and Fig.~\ref{fig:fc_10_comparison} show the convergence behavior with different learning rates $\eta \in \{0.05, 0.1, 0.3\}$. Increasing the value of $\eta$ allow the algorithm to take bigger steps and to converge faster in case of two nodes.  
But, when $N=10$, the higher learning rate brings more fluctuations due to increased noise in the utility estimates.
The evaluation with varying exploration parameters was skipped due to the space limit.

\subsection{Heterogeneous Traffic}

Here, we evaluate the performance of the algorithm under heterogeneous traffic, where the optimal $CW$ values are not identical for all transmitting nodes. We consider two scenarios, where three nodes use diverse transmission parameters: \textit{i)} the same data rate (MCS3, 26\,Mbps), but diverse packet sizes $D_i \in \{250, 500, 1000\}$\,B; \textit{ii)} the same packet size (1500\,B), but diverse data rates $M_i \in \{6.5, 26, 65\}$\,Mbps.

\begin{figure}[ht!]
\centering
  \includegraphics[width=1.0\linewidth]{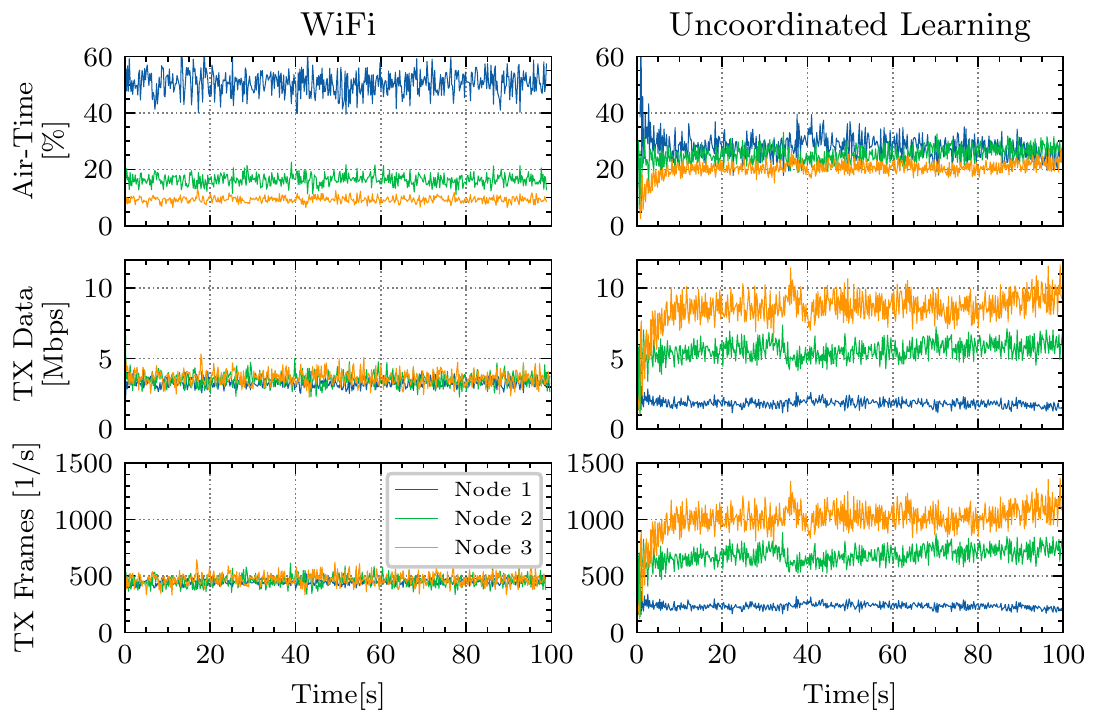}
  \vspace{-23pt}
  \caption{Air-time allocation, individual throughput and packet rate in case of heterogenous traffic, i.e., all nodes use the same packet size of 1500\,B, but different data rates $M_i \in \{6.5, 26, 65\}$\,Mbps.}
  \label{fig:fc_hetero_mcs_comparison}
\end{figure}

\begin{figure}[ht!]
\centering
  \includegraphics[width=1.0\linewidth]{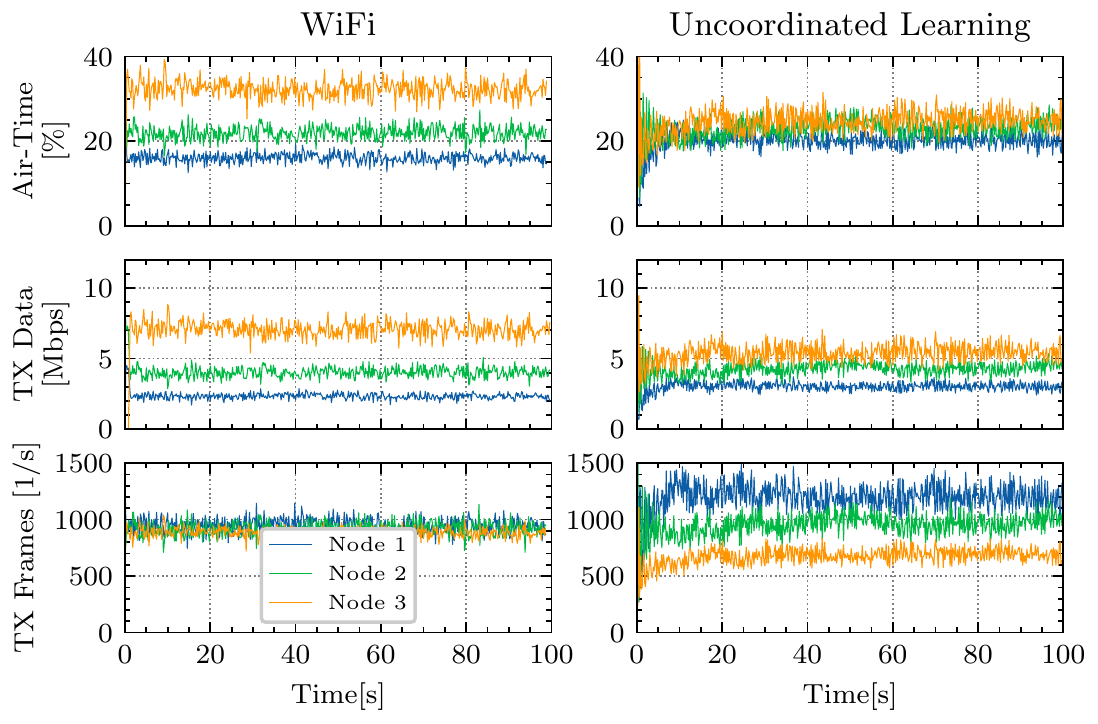}
  \vspace{-20pt}
  \caption{Air-time allocation, individual throughput and packet rate in case of heterogenous traffic, i.e., all nodes use the same data rate of  26\,Mbps (MCS3), but different packet sizes $D_i \in \{250, 500, 1000\}$\,B.} 
  \label{fig:fc_hetero_pkt_comparison}
  \vspace{-10pt}
\end{figure}

In Fig.~\ref{fig:fc_hetero_mcs_comparison} and Fig.~\ref{fig:fc_hetero_pkt_comparison}, we compare the performance of our distributed approach with standard WiFi operation. Specifically, we are interested in air-time allocation, individual throughputs and packets data rate.
Due to the symmetric contention process, WiFi assures an equal number of transmission opportunities to all nodes, i.e., frame-fairness.
Our results show the \textit{performance anomaly} problem in WiFi, i.e., despite using the higher data rate, the performance of the faster nodes is capped at that of the slowest station.
In contrast, the proposed algorithm allows nodes to successfully and quickly converge to equal air-time allocation (i.e., proportional-fair allocation) in both scenarios.

\subsection{Dynamic Scenario}

\begin{figure}[ht!]
\centering
  \includegraphics[width=1.0\linewidth]{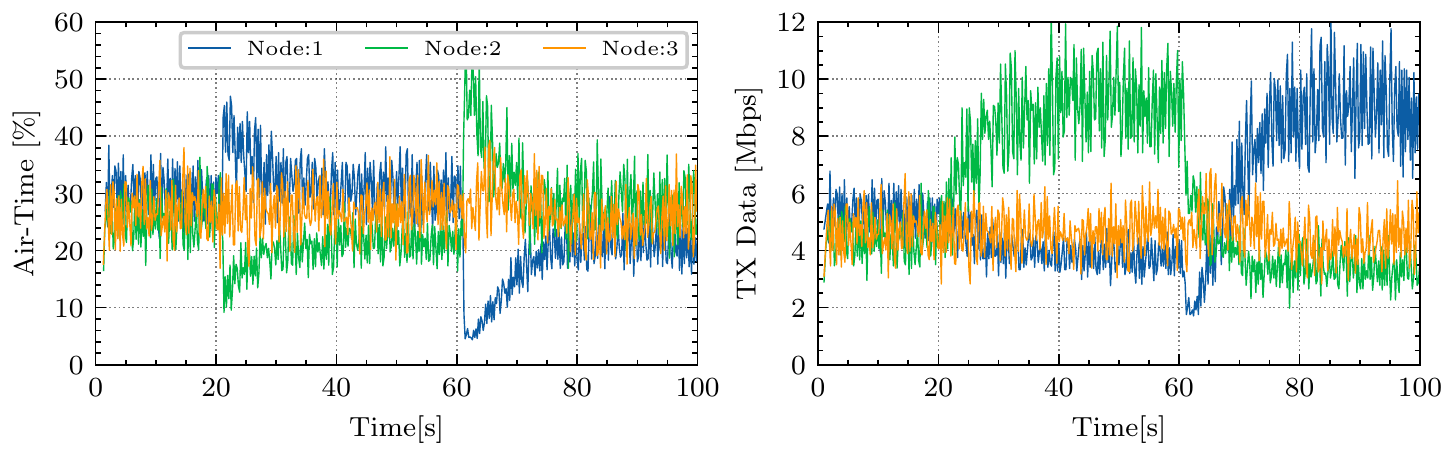}
  \vspace{-12pt}
  \caption{Air-time allocation and individual throughput in dynamic scenario. Nodes starts with the same data rate. At $t=20$\,s and $t=60$\,s, node 1 and 2 change the used data rate.} 
  \label{fig:dynamic_scenario}
  \vspace{-12pt}
\end{figure}

Using a setup similar as in the previous sections, we evaluate the adaptability of the proposed algorithm to network dynamics.
Specifically, we consider a dynamic scenario with three transmitters that change data rates (e.g., due to the change of wireless propagation condition).
Fig.~\ref{fig:dynamic_scenario} show the individual air-time allocation and throughput. The traffic is saturated, and nodes use a packet size of 1000\,B. The nodes starts with the same data rate of 13\,Mbps (MCS2), then at time  $t=20$\,s, nodes change the data rates, i.e., Node-1 switches to 65\,Mbps (MCS7) and Node-2 switches to 6.5\,Mbps (MCS0). At $t=60$\,s, nodes change data rates again, i.e., Node-1 switches to 6.5\,Mbps (MCS0) and Node-2 switches to 65\,Mbps (MCS7). Node-3 never changes its data rate.

We observe in Fig.~\ref{fig:dynamic_scenario} that the algorithm can adapt to the changes in data rates. The convergence takes around 10\,s after the change occurs.

\subsection{Unsaturated traffic}

Here, we evaluate the behavior of the proposed algorithm under unsaturated traffic conditions. Specifically, we simulate two scenarios with three transmitting nodes, in which: \textit{i)} the total offered load does not saturate the wireless channel, i.e., $r_i \in \{200, 400, 600\}$\,pkts/s; \textit{ii)} the total offered load saturates the wireless channel as one node operates with saturated traffic, i.e., $r_i \in \{200, 400, 2000\}$\,pkts/s.
The nodes use homogeneous transmission parameters, namely a data rate of 26\,Mbps (MCS3), and a packet size of 1000\,B.

\begin{figure}[t!]
  \centering
  \includegraphics[width=\linewidth]{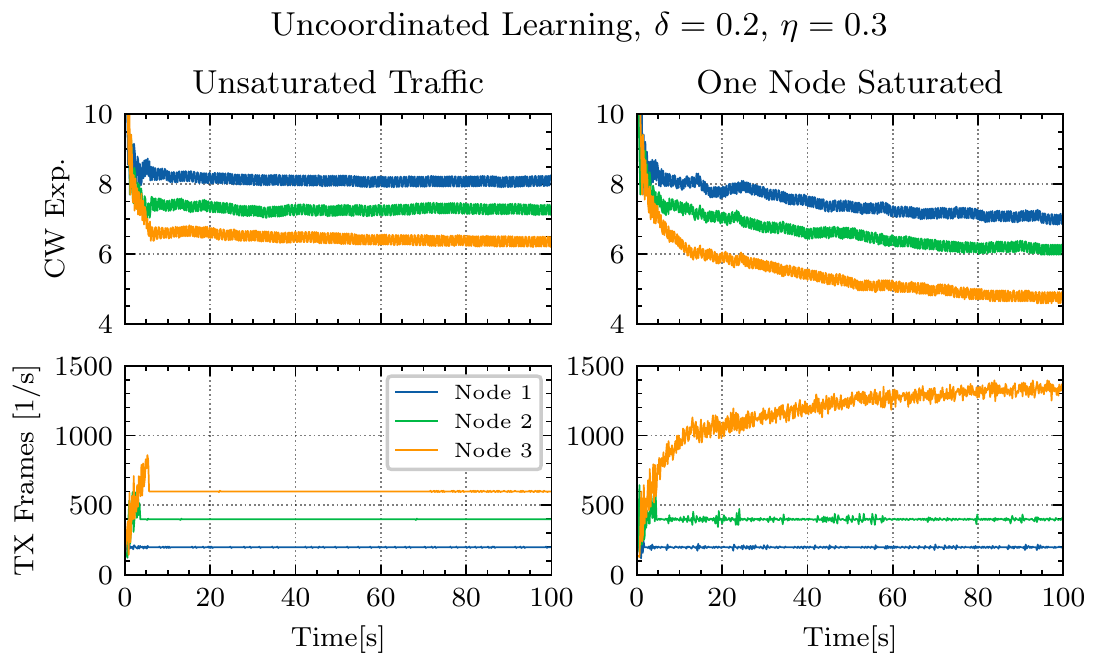}
  \vspace{-20pt}
  \caption{Individual contention window and packet rate in unsaturated scenario (left) with offered load of  $r_i \in \{200, 400, 600\}$\,pkts/s and with one node saturated (right), i.e., $r_i \in \{200, 400, 2000\}$\,pkts/s.} 
  \label{fig:nonsat}
  \vspace{-10pt}
\end{figure}

Fig.~\ref{fig:nonsat} shows the evolution of individual contention window and packet rate in both scenarios. We observe that in the case of unsaturated traffic the nodes increase their aggressiveness until the offered load is satisfied. Afterward, they operate with a stable $CW$, i.e., the perturbation of the $CW$ does not change the value of the network utility, and hence the estimated gradient equals zero. 
In the second scenario, the node with a high traffic load increases its aggressiveness until it saturates the wireless channel, but without negatively affecting the slower nodes, i.e., they also adapt their CW properly to get enough transmission opportunities and satisfy their own traffic.

\subsection{Flow-In-the-Middle Topology}

\begin{figure}[ht!]
\centering
    \includegraphics[width=0.4\linewidth]{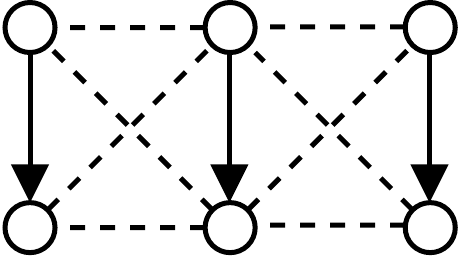}
    \caption{Flow-In-the-Middle (FIM) Topology. In the network graph, vertices, dotted lines, and arrows represent nodes,  connectivity, and flows, respectively.}  
    \label{fig:fitm_topology}
    \vspace{-12pt}
\end{figure}

Finally, we evaluate the behavior of the proposed scheme in a flow-in-the-middle (FIM) topology --  Fig.~\ref{fig:fitm_topology}. 
The FIM topology is a simple multi-collision domain scenario, where nodes have asymmetric contention information.
Specifically, the central transmitter can carrier sense transmissions of both its neighbors, while the edge transmitters cannot carrier sense each other. Therefore, the middle transmitter defers its transmissions whenever at least one of its neighbors transmits a frame, while concurrent transmissions of the edge nodes can occur. Note that the transmissions of the edge nodes may interleave, leaving no silent periods for the middle node. As a result, the throughput of the middle node is lowered due to the lack of transmission opportunities. In the worst case, the middle node suffers from complete starvation~\cite{aryafar2013csma}.

We simulate a scenario where all three transmitters use the same data rate of 26\,Mbps (MCS3), and packet size of 1500\,B. The traffic is saturated. 
In the FIM topology, only the middle transmitter can estimate the global utility function, while the edge nodes can estimate the utility function only in their collision domain. We consider two cases of $CW$ learning, namely uncoordinated learning, where nodes locally estimate the utility function, and coordinated learning, where nodes perform gradient estimation procedure synchronously and the global function computed by the middle node is communicated to the edge nodes. 

\begin{figure}[ht!]
  \centering
  \includegraphics[width=\linewidth]{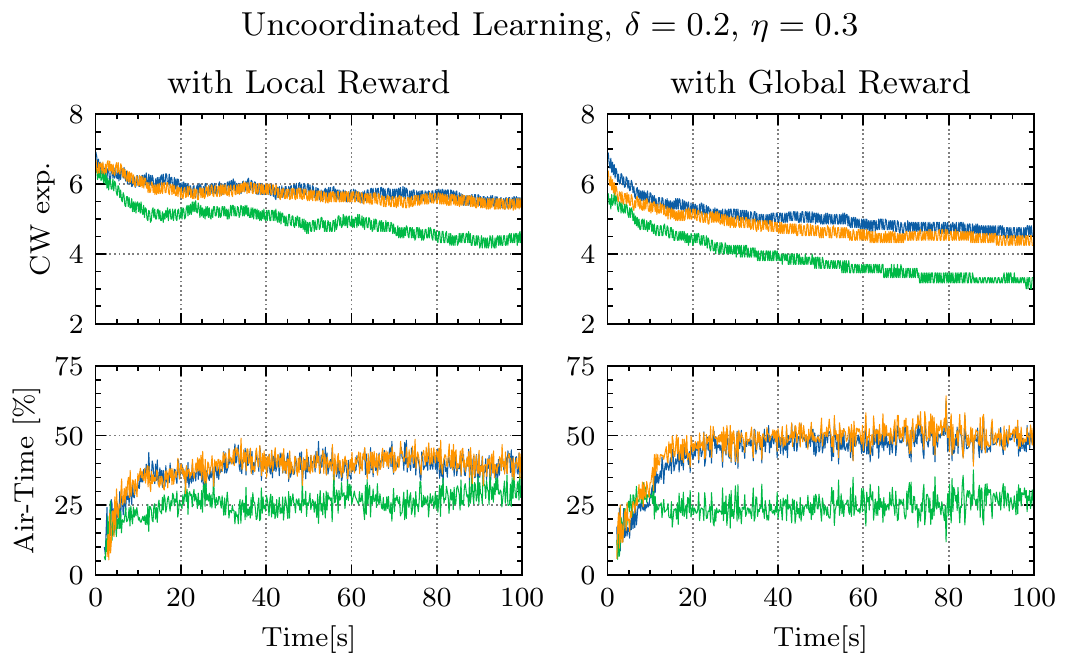}
  \vspace{-20pt}
  \caption{The individual contention window and air-time allocation in FIM topology for uncoordinated learning with locally estimated utility (left) and coordinated learning with global utility (right).} 
  \label{fig:fitm_comparison}
  \vspace{-10pt}
\end{figure}

\begin{figure}[ht!]
  \centering
  \includegraphics[width=\linewidth]{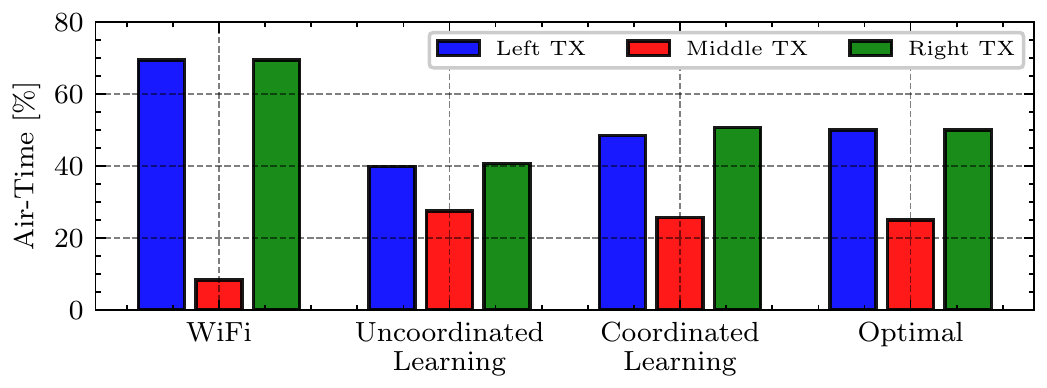}
  \vspace{-20pt}
  \caption{The averaged air-time allocation attained by 802.11 WiFi and using the proposed learning-based approach in the FIM topology. Optimal air-time allocation is presented for comparison.} 
  \label{fig:fitm_comparison_air_time}
  \vspace{-10pt}
\end{figure}

Fig.~\ref{fig:fitm_comparison} shows the evolution of the individual $CW$ and the air-time allocation, while the Fig.~\ref{fig:fitm_comparison_air_time} show the air-time allocation averaged over the simulation duration (i.e., 100\,s).
Our results confirm the problem of starvation of the middle node in the case of standard WiFi BEB operation. Specifically, the middle node gets only 9\% of the channel air-time, while the edge nodes around 70\%. We also show the optimal air-time allocation found with extensive simulations.
The proposed algorithm improves the fairness among nodes, i.e., the middle node gets assigned more air-time while the edge nodes get proportionally less. When using the global utility, the nodes can find the $CW$ values leading to the optimal solution. However, with local utilities, the middle node becomes too aggressive. Moreover, as the goals of the three nodes are not consistent, the distributed algorithm cannot converge. Instead, the algorithm oscillates between two solutions, i.e., the middle node wants to achieve proportional fairness for three nodes, while the edge nodes optimize its operation for two nodes.

%% file: sections/conclusions.tex

\section{Conclusions}

In this paper, we apply a distributed Kiefer-Wolfowitz algorithm to a wireless network optimization problem. 
We address the distributed tuning of contention windows in WiFi networks and show that nodes can learn the proper contention windows values yielding proportional fair resource allocation even without explicit communication. Specifically, they can estimate the global utility function by overhearing each others' transmissions and use it to solve the log-convex optimization problem collaboratively.
We evaluated the algorithm in multiple scenarios and with various levels of coordination between nodes (i.e., time and action synchronization). 
We conclude that based on the KW algorithm, one can design simple yet powerful learning-based collaboration schemes, which do not require any communication nor coordination among agents. 
However, the parameters of such an algorithm have to be carefully configured to make the convergence speed practical. The automatic tuning of the parameters is left for further study.

%% file: main.bbl
\begin{thebibliography}{10}
\providecommand{\url}[1]{#1}
\csname url@samestyle\endcsname
\providecommand{\newblock}{\relax}
\providecommand{\bibinfo}[2]{#2}
\providecommand{\BIBentrySTDinterwordspacing}{\spaceskip=0pt\relax}
\providecommand{\BIBentryALTinterwordstretchfactor}{4}
\providecommand{\BIBentryALTinterwordspacing}{\spaceskip=\fontdimen2\font plus
\BIBentryALTinterwordstretchfactor\fontdimen3\font minus
  \fontdimen4\font\relax}
\providecommand{\BIBforeignlanguage}[2]{{%
\expandafter\ifx\csname l@#1\endcsname\relax
\typeout{** WARNING: IEEEtran.bst: No hyphenation pattern has been}%
\typeout{** loaded for the language `#1'. Using the pattern for}%
\typeout{** the default language instead.}%
\else
\language=\csname l@#1\endcsname
\fi
#2}}
\providecommand{\BIBdecl}{\relax}
\BIBdecl

\bibitem{1208921}
M.~{Heusse}, F.~{Rousseau}, G.~{Berger-Sabbatel}, and A.~{Duda}, ``Performance
  anomaly of 802.11b,'' in \emph{IEEE INFOCOM}, 2003.

\bibitem{patras2016proportional}
P.~Patras, A.~Garcia-Saavedra, D.~Malone, and D.~J. Leith, ``Rigorous and
  practical proportional-fair allocation for multi-rate wi-fi,'' \emph{Ad Hoc
  Networks}, 2016.

\bibitem{kelly1998proportional}
F.~Kelly, A.~Maulloo, and D.~Tan, ``Rate control in communication networks:
  shadow prices, proportional fairness and stability,'' \emph{Journal of the
  Operational Research Society}, 1998.

\bibitem{checco2011fairness}
A.~{Checco} and D.~J. {Leith}, ``{Proportional Fairness in 802.11 Wireless
  LANs},'' \emph{IEEE Communications Letters}, 2011.

\bibitem{5370273}
D.~J. {Leith}, V.~G. {Subramanian}, and K.~R. {Duffy}, ``{Log-convexity of rate
  region in 802.11e WLANs},'' \emph{IEEE Communications Letters}, 2010.

\bibitem{walrand2020distributed}
J.~{Walrand}, ``{Convergence of a Distributed Kiefer-Wolfowitz Algorithm},''
  \emph{arXiv e-prints}, p. arXiv:2008.12856, Aug. 2020.

\bibitem{kiefer1952stochastic}
J.~Kiefer and J.~Wolfowitz, ``{Stochastic estimation of the maximum of a
  regression function},'' \emph{The Annals of Mathematical Statistics}, 1952.

\bibitem{6195728}
{Yuan Le}, {Liran Ma}, {Wei Cheng}, {Xiuzhen Cheng}, and {Biao Chen},
  ``Maximizing throughput when achieving time fairness in multi-rate wireless
  lans,'' in \emph{IEEE INFOCOM}, 2012.

\bibitem{6598666}
Y.~{Le}, L.~{Ma}, W.~{Cheng}, X.~{Cheng}, and B.~{Chen}, ``{A Time
  Fairness-Based MAC Algorithm for Throughput Maximization in 802.11
  Networks},'' \emph{IEEE Transactions on Computers}, 2015.

\bibitem{5463215}
M.~{Laddomada}, F.~{Mesiti}, M.~{Mondin}, and F.~{Daneshgaran}, ``{On the
  throughput performance of multirate IEEE 802.11 networks with variable-loaded
  stations: analysis, modeling, and a novel proportional fairness criterion},''
  \emph{IEEE Transactions on Wireless Comm.}, 2010.

\bibitem{Golshan2020bco}
G.~Famitafreshi and C.~Cano, ``{Achieving Proportional Fairness in WiFi
  Networks via Bandit Convex Optimization},'' in \emph{{Machine Learning for
  Networking}}.\hskip 1em plus 0.5em minus 0.4em\relax Springer International
  Publishing, 2020.

\bibitem{bianchi2000performance}
G.~{Bianchi}, ``{Performance analysis of the IEEE 802.11 distributed
  coordination function},'' \emph{IEEE Journal on Selected Areas in Com.},
  2000.

\bibitem{spall1992spsa}
J.~C. {Spall}, ``{Multivariate stochastic approximation using a simultaneous
  perturbation gradient approximation},'' \emph{IEEE Transactions on Automatic
  Control}, 1992.

\bibitem{ns3}
{ns-3 network simulator}, \textit{https:\/\/www.nsnam.org\/}, accessed:
  2021-02-05.

\bibitem{ns3gym}
P.~Gaw{\l}owicz and A.~Zubow, ``{ns-3 meets OpenAI Gym: The Playground for
  Machine Learning in Networking Research},'' in \emph{{ACM MSWiM}}, 2019.

\bibitem{aryafar2013csma}
E.~{Aryafar}, T.~{Salonidis}, J.~{Shi}, and E.~{Knightly}, ``{Synchronized CSMA
  Contention: Model, Implementation, and Evaluation},'' \emph{IEEE/ACM
  Transactions on Networking}, 2013.

\end{thebibliography}
